\documentstyle[12pt,aasms4]{article}

\lefthead{Bekki Kenji and   Yasuhiro Shioya}
\righthead{Formation and evolution of dusty starburst galaxies}

\begin{document}
\title{Formation and evolution of dusty starburst galaxies I.
A new method for deriving spectral energy distribution}

\author{Kenji Bekki} 
\affil{Division of Theoretical Astrophysics,
National Astronomical Observatory, Mitaka, Tokyo, 181-8588, Japan} 

\and

\author{Yasuhiro Shioya} 
\affil{
Astronomical Institute, 
Tohoku University, Sendai, 980-8578, Japan}

\begin{abstract}

We present a new numerical code which is designed to derive a spectral
energy distribution (SED) for 
an arbitrary spatial distribution of stellar and gaseous components
in a dusty starburst galaxy.
We apply a  ray tracing method to  numerical simulations
and thereby  estimate extinction and  reemission of stellar light
by dusty gas in an explicitly self-consistent manner. 
By using this code, we can investigate simultaneously dynamical and 
photometric evolution of a dusty galaxy
based on stellar and gaseous dynamical simulations.
As an example, we demonstrate when and how a major galaxy
merger with dusty starburst 
becomes an ultra-luminous infrared galaxy owing to strong internal
dust extinction.
We furthermore discuss advantages and disadvantages of the present
new code in clarifying the nature and the origin of low and high
redshift dusty starburst galaxies.

\end{abstract}


\section{Introduction}

  Recent observational studies have discovered
possible dusty starburst candidates in various classes of
galactic objects such as 
low z ultra-luminous infrared galaxies (ULIRGs)
(e.g., Sanders et al. 1988;  Sanders \& Mirabel 1996),
intermediate z ones (e.g., Tran et al. 1999),
faint submillimeter sources detected
by  the Sub-millimeter Common-User Bolometer Array
(SCUBA) (Holland et al. 1999) on the James Clerk Maxwell Telescope
(Smail, Ivison, \& Blain 1997; 
Hughes et al. 1998; Barger et al. 1998;
Smail et al. 1998; Ivison et al. 1999; Lilly et al.  1999),
extremely red objects (EROs)
 (Elston, Rieke, \& Rieke 1988; Graham \& Dey 1996;
Cimatti et al. 1998;
Dey et al. 1999; Smail  et al. 1999),
optically faint radio sources detected by the Very
Large Array (VLA) (e.g., Richards et al. 1999),
Lyman-break galaxies (Steidel et al. 1996; Lowenthal et al. 1997),
and near-infrared emission line galaxies (e.g., Mannucci et al. 1998). 
These dusty starburst candidates are generally considered to provide valuable
information on the formation and the evolution of galaxies,
accordingly the origin and the nature of these galaxies
have been extensively discussed in variously different contexts.
These discussions include, for example,
 the importance of dissipative dynamics
in the formation of elliptical galaxies at low and high redshift
(Kormendy \& Sanders 1992; Barnes \& Hernquist 1992),
an evolutionary link between starburst and active galactic nuclei
(Norman \& Scoville 1988), fueling of dusty gas to the central
$\sim$ 100 pc for nuclear
starburst in major galaxy mergers (e.g., Mihos \& Hernquist 1996),
physical correlations between morphological and photometric
properties in ULIRGs (Bekki, Shioya, \& Tanaka 1999; Bekki \& Shioya 2000a),
merging and clustering process of high $z$ dusty mergers
within a hierarchical clustering scenario
(e.g., Somerville \& Primack 1998),
and cosmic star formation history (e.g., Pei \& Fall 1995;
Madau et al. 1996; Meurer et al. 1997; Madau, Pozzetti, \& Dickinson 1998;
Pascarelle, Lanzetta, \& Fernandez-Soto 1998;
Blain et al. 1999; 
Steidel et al. 1999).
One of important problems in extra-galactic astronomy
is to settle the above discussions on high redshift  dusty 
starburst galaxies, 
though observational analysis
still has some
difficulties in determining  precisely  redshifts (Sanders 1999),
identifying  optical counterparts (Richards 1999), and
estimating the degree of dust extinction (e.g., 
Meurer, Heckman, \& Calzetti 1999)
for high $z$ dusty starburst galaxies.

Spectral energy distributions  (SEDs) are generally considered
to be one of essentially important factors that can determine
the nature of dusty starburst galaxies. 
Accordingly several theoretical attempts have been made to 
derive the SEDs of galaxies with dusty starburst regions.
Rowan-Robinson \& Crawford (1989) discussed how the far-infrared 
spectra of a sample of galaxies 
observed by $Infrared$ $Astronomical$ $Satellite$ (IRAS) 
can be reproduced
by a theoretical model with  a given
temperature of stars embedded by dust,
the optical depth,
and the density distribution (or geometry) of dust.   
Witt et al. (1991) investigated 
transfer processes of stellar light for models with variously
different spherical geometries of dust 
with  special emphasis on 
the effects of scattered light  on photometric properties
of galaxies 
and SEDs. 
Calzetti, Kinney, \& Storchi-Bergmann (1994)
analyzed ultraviolet (UV) and optical spectra of 39 starburst and blue
compact galaxies and thereby provided
an analytical formulation of the effects of dust extinction
in galaxies and an effective extinction law for correcting
the observed UV and optical spectral continua.
Witt \& Gordon (1996) investigated the radiative transfer
processes of a central stellar source surrounded by a spherical,
statistically homogeneous but clumpy two-phase scattering medium
and found that the structure 
of dusty medium 
can greatly affect the conversion of UV, optical, and near-infrared radiation
into thermal far-IR dust radiation in a dusty system.
By comparing the observed SEDs of 30 starburst galaxies with
theoretical radiative transfer models  of dusty systems,
Gordon, Calzetti, and Witt (1997) 
discussed the importance of geometry of stellar and gaseous components
in determining the SED of a dusty starburst galaxy.
Takagi, Arimoto, \& Vansevi$\rm \check{c}$ius (1999) investigated
radiative transfer models with variously different 
ages of secondary starburst components and optical depths
for dusty starburst galaxies and proposed 
a new method for estimating  precisely 
the effect of ages 
of young starburst populations  and that of dust attenuation 
on the shape of SED in UV and near$-$infrared bands.  
Efstathiou, Rowan-Robinson, \& Siebenmorgen (2000)
treated starburst galaxies as an ensemble of optically
thick giant molecular clouds (GMCs)
centrally illuminated by recently formed stars
and thereby constructed a new radiative transfer model
for calculating SEDs from UV to millimeter band of dusty
starburst galaxies.
By using this model, they discussed how the age and  
the star formation history of a dusty starburst galaxy
can control the SED
particularly for the prototypical starburst galaxy M82
and NGC 6090.

Although the above previous studies have succeeded
in clarifying  
important dependences of SEDs on physical parameters
such as spatial distribution 
of dust, geometries of stellar and gaseous components,
and dust properties
in  dusty starburst galaxies, 
many of them  did not discuss $the$ $time$ $evolution$ $of$ $SEDs$.
Main reasons for $some$ previous theoretical
studies' not discussing the time evolution of SEDs
are the following three.
Firstly previous models did not follow the time
evolution of stellar and gaseous distribution
and accordingly could not derive the time evolution
of SEDs in dusty starburst galaxies.
Secondly,  hydro-dynamical evolution of interstellar gas
(e.g., time evolution of gaseous density)
was not included in previous studies,
 and accordingly the time evolution of optical depth
could  not be derived.
Thirdly, some previous studies 
did not consider chemical evolution  of  dusty interstellar medium 
and that of stellar components,
they could not follow  time evolution of metallicity
and that of dust properties.
Considering that low z infrared luminous galaxies with
dusty starburst and high $z$ faint SCUBA sources with
possible dusty starburst show very peculiar morphology 
(Sanders et al. 1988;  Sanders \& Mirabel 1996; Smail et al. 1998),
spherical symmetric approximation  or   axisymmetric one 
adopted in previous theoretical and numerical studies
for  the stellar and gaseous distribution of a dusty galaxy
should be also relaxed in order that the nature and the origin 
of low and high z dusty galaxies can be  extensively investigated
by theoretical studies.

The purpose of the present paper and our future papers
is to investigate in detail the formation
and the evolution of dusty starburst galaxies
by using a new numerical code for deriving the time evolution
of SEDs of these galaxies.
We first perform numerical simulations that can follow dynamical evolution
of stellar and gaseous components, star formation history,
and chemical evolution for dusty starburst 
galaxies in an explicitly self-consistent manner.
We then derive stellar and gaseous distribution 
and age and metallicity distribution of stellar populations
and thereby calculate galactic SEDs.
Furthermore  we  describe
how the present  numerical code is useful and helpful
for investigating variously different physical properties
of low and high z dusty galaxies. 
As an example,  we here 
present the results of a major merger with dusty interstellar gas.
We particularly demonstrate how dynamical evolution of stellar and gaseous
components controls the time evolution of SEDs in
dusty galaxy mergers  and thus their photometric
evolution.

The layout of this paper is as follows.
In \S    2, we summarize  numerical models used in the
present study and describe   in detail the methods for 
deriving the SEDs corrected by internal dust extinction.
In this section, we also point out
the limitations of the present numerical code.
In \S 3, we present numerical results on the time evolution
of morphology, SED, and photometric properties in a gas-rich
major merger. In \S 4, we discuss the origin of high $z$ 
dusty starburst galaxies such as faint SCUBA sources and EROs. 
The conclusions of the preset study
are  given in \S 5.

\section{Model}

The most remarkable difference in deriving SEDs 
of dusty galaxies between the present model and previous  models
(e.g., Mazzei, Xu, \& De Zotti 1992; Franceschini et al. 1994;
Witt \& Gordon 1996;  Gordon  st al. 1997;
Guiderdoni et al. 1998) 
is that we derive SEDs of galaxies
based on the result of numerical simulations that can follow 
both  dynamical and chemical evolution of galaxies.
The derivation of a SED for a galaxy with dusty starburst
consists of the following three steps.
In the first step we simultaneously derive age and metallicity
distribution of stellar component in the galaxy, based on
numerical simulations.
Secondly, we use a stellar population synthesis code (e.g.
Bruzal \& Charlot 1993) and thereby calculate the SED of the galaxy,
based on the derived age and metallicity distribution of
stellar population in the galaxy. 
In this second step, the dust effects are not included.
Thirdly, we consider  extinction and  reemission of stellar
light by dusty interstellar gas and modify the SED of the galaxy,
by using our new code.
The details of our new  method for modifying  galactic SEDs
are given in \S 2.3.

\subsection{Merger  model}

\subsubsection{Initial conditions}

 We construct  models of galaxy mergers between gas-rich 
 disk galaxies with equal mass by using Fall-Efstathiou model (1980).
 The total mass and the size of a progenitor disk are $M_{\rm d}$
 and $R_{\rm d}$, respectively. 
 From now on, all the mass and length are measured in units of
  $M_{\rm d}$ and  $R_{\rm d}$, respectively, unless specified. 
  Velocity and time are 
  measured in units of $v$ = $ (GM_{\rm d}/R_{\rm d})^{1/2}$ and
  $t_{\rm dyn}$ = $(R_{\rm d}^{3}/GM_{\rm d})^{1/2}$, respectively,
  where $G$ is the gravitational constant and assumed to be 1.0
  in the present study. 
  If we adopt $M_{\rm d}$ = 6.0 $\times$ $10^{10}$ $ \rm M_{\odot}$ and
  $R_{\rm d}$ = 17.5 kpc as a fiducial value, then $v$ = 1.21 $\times$
  $10^{2}$ km/s  and  $t_{\rm dyn}$ = 1.41 $\times$ $10^{8}$ yr,
  respectively.
  In the present model, the rotation curve becomes nearly flat
  at  0.35  radius with the maximum rotational velocity $v_{\rm m}$ = 1.8 in
  our units.
  The corresponding total mass $M_{\rm t}$ and halo mass $M_{\rm h}$
  are 5.0  and 4.0 in our units, respectively,
which means that the baryonic mass fraction of the initial disk galaxy
is 0.2.
  The radial ($R$) and vertical ($Z$) density profile 
  of a  disk are  assumed to be
  proportional to $\exp (-R/R_{0}) $ with scale length $R_{0}$ = 0.2
  and to  ${\rm sech}^2 (Z/Z_{0})$ with scale length $Z_{0}$ = 0.04
  in our units,
  respectively.
  The velocity dispersion
  of halo component 
   at a given point
  is set to be isotropic and given
  according to the  virial theorem.
  In addition to the rotational velocity made by the gravitational
  field of disk and halo component, the initial radial and azimuthal velocity
  dispersion are given to disk component according
  to the epicyclic theory with Toomre's parameter (\cite{bt87}) $Q$ = 1.2.
  The vertical velocity dispersion at given radius 
  is set to be 0.5 times as large as
  the radial velocity dispersion at that point, 
  as is consistent  with 
  the observed trend  of the Milky Way (e.g., Wielen 1977).
 As is described above, the present initial disk model does not
 include any remarkable bulge components, and accordingly corresponds to
 `purely'  late-type spiral without galactic bulge. Although it is
 highly possible that  galactic bulges greatly affect the chemical evolution
 of galaxy mergers, we however investigate
 this issue in our future papers.

  The collisional and dissipative nature 
  of the interstellar medium is  modeled by the sticky particle method
  (\cite{sch81}).
It should be emphasized here that this discrete cloud model can at best represent
the $real$ interstellar medium of galaxies  in a schematic way. 
As is modeled by McKee \& Ostriker (1977),
the interstellar medium can be considered to be  
 composed mainly of `hot', `warm', and `cool'
gas,
each of which mutually
interacts hydrodynamically 
 in a rather  complicated way.
 Actually, these considerably complicated nature of
interstellar medium in  disk galaxies would not be
  so simply modeled by the `sticky
particle' method in which gaseous dissipation is modeled by ad hoc
cloud-cloud collision: Any existing numerical method probably could
not model the $real$ interstellar medium in an admittedly proper
way. 
In the present study, as a compromise,
we only try to address some important aspects of hydrodynamical
interaction between interstellar medium in disk galaxies and in
dissipative mergers. 
More elaborated numerical modeling for real interstellar medium
would be  necessary for 
our further understanding of dynamical evolution 
in dissipative galaxy mergers. 
  We assume that the fraction of gas mass ($f_{g}$) in
  a disk is set to be 0.5 initially.
Actually, the gas mass fraction in precursor disks of a merger
is  different between galaxy mergers  and depends on the epoch of
the merging.
For example, recent observational results on the total mass
of molecular gas for faint SCUBA sources (e.g., Frayer et al. 1999) 
revealed 
that SMM J14011+0252 with some indications of major merging
at z = 2.565 has $\sim$ 5.0 $\times$ $10^{10}$ $M_{\odot}$.
If the progenitor of this gas-rich high redshift merger has a mass of      
6.0 $\times$ $10^{10}$ $ \rm M_{\odot}$ (i.e., the same as that of the Galaxy),
the gas mass fraction is $\sim$ 0.42 for this galaxy.
The gas mass fraction is observationally suggested
to be larger for higher redshift mergers (e.g., Evans, Surace, \&
Mazzarella 1999).
Guided by these observational results, we adopt the above value
(0.5) as the initial gas mass fraction in order to discuss
higher redshift galaxy mergers. 
Although the  difference of gas mass fraction
probably could yield a great variety of chemical
and dynamical structures in mergers, we do not intend to
consider this important difference for simplicity in the present  paper
and will address in our future paper. 
  The radial and tangential restitution coefficient for cloud-cloud
  collisions are
  set to be 1.0 and
  0.0, respectively.
  The  number of particles 
  for an above  isolated galaxy is 
  10000 for dark halo,   
10000 for stellar disk components, and 10000 for gaseous ones.

    In all of the simulations of mergers, the orbit of the two disks is set to be
    initially in the $xy$ plane and the distance between
    the center of mass of the two disks,
represented by $r_{\rm in}$,
  is  assumed  to be the  free parameter 
which controls the epoch of galaxy merging.
    The pericenter
    distance, represented by $r_{\rm p}$, is also
assumed to be the  free parameter which controls the initial
total orbital  angular momentum of galaxy mergers.
The eccentricity
is  set to be 1.0 for all models of  mergers,
meaning that the encounter of galaxy merging is parabolic.
    The spin of each galaxy in a  pair merger
is specified by two angle $\theta_{i}$ and
    $\phi_{i}$, where suffix  $i$ is used to identify each galaxy.
    $\theta_{i}$ is the angle between the $z$ axis and the vector of
    the angular momentum of a disk.
    $\phi_{i}$ is the azimuthal angle measured from $x$ axis to
     the projection of the angular momentum vector of a disk onto
     $x$-$y$ plane. 
In the present study, we show the results of only one
model with $\theta_{1}$ = 0.0, $\theta_{2}$ = $-150.0$,
$\phi_{1}$ = 0.0,  $\phi_{2}$ = 0.0, $r_{\rm p}$ = 17.5 kpc,
and  $r_{\rm in}$ = 140 kpc: This model describes a nearly 
prograde-retrograde merger.
The results of the models with variously
different  $\theta_{i}$, $\phi_{i}$, $r_{\rm p}$ 
and $r_{\rm in}$ will be  described in our future papers. 
The time when the progenitor disks merge completely and reach  the
dynamical equilibrium is less than 15.0 in our units for most of
models and does not depend so
strongly on the  history of star formation in the  present calculations.

\subsubsection{Global star formation}

    Star formation
     is modeled by converting  the collisional
    gas particles
    into  collisionless new stellar particles according to the algorithm
    of star formation  described below.
    We adopt the Schmidt law (Schmidt 1959)
    with exponent $\gamma$ = 2.0 (1.0  $ < $  $\gamma$
      $ < $ 2.0, \cite{ken89}) as the controlling
    parameter of the rate of star formation.
    The amount of gas 
    consumed by star formation for each gas particle
    in each time step, 
    $\dot{M_{\rm g}}$, 
is given as:
    \begin{equation}
      \dot{M_{\rm g}} \propto  C_{\rm SF} \times 
 {(\rho_{\rm g}/{\rho_{0}})}^{\gamma - 1.0}
    \end{equation}
    where $\rho_{\rm g}$ and $\rho_{0}$
    are the gas density around each gas particle and
    the mean gas density at 0.48 radius  of 
    an initial disk, respectively.
    This density-dependent star formation model 
is similar to that of Mihos, Richstone, \& Bothun  (1992)
and the probability approach of the gas consumption rate
is similar to that described by Katz (1992) and O'Neil et al. (1998). 
    In order to avoid a large number of new stellar particles with
different mass, we convert one gas particle into one stellar one
according to the following procedure.
First we give each gas particle the probability, $P_{\rm sf}$,
 that the gas particle
is converted into stellar one, by setting the $P_{\rm sf}$ to be proportional
to the  $\dot{M_{\rm g}}$  in equation (1) estimated for the gas particle. 
Then we draw the random number to determine whether or not the gas particle
is totally converted  into one new star. 
This method of star formation enables us to control the rapidity of star formation
without increase of particle number in each simulation thus to maintain
the numerical accuracy in  each simulation. 
    The $C_{\rm SF}$ in the equation (1)
is the parameter that controls the rapidity of 
    gas consumption by star formation.
We determine the initial   value of  the $C_{\rm SF}$ 
so  that mean star formation rate of an isolated late-type disk
galaxy model with the typical
gas mass fraction of $\sim$ 0.2 
can become an order of 1 $ M_{\odot}$ ${\rm yr}^{-1}$ for the first 1 Gyr
evolution.
The positions and velocity of the new stellar particles are set to 
be the same as those of original gas particles.

   All the calculations related to 
the above dynamical evolution  including the dissipative
dynamics, star formation, and gravitational interaction between collisionless
and collisional component 
 have been carried out on the GRAPE board
   (\cite{sug90})
   at Astronomical Institute of Tohoku University.
   The parameter of gravitational softening is set to be fixed at 0.03  
   in all the simulations. The time integration of the equation of motion
   is performed by using 2-order
   leap-flog method. Energy and angular momentum  are conserved
within 1 percent accuracy in a test collisionless merger simulation.
Most of the  calculations are set to be stopped at T = 20.0 in our units
unless specified.

\subsection{Method for the SED of stellar populations}

\subsubsection{Chemical enrichment}

 Chemical enrichment through star formation during galaxy merging
is assumed to proceed both locally and instantaneously in the present study.
The model for analyzing
metal enrichment of each gas  and stellar particle 
is as follows.
First, as soon as a gas particle is converted into a new stellar one by
star formation, we search neighbor gas particles locating within
$R_{\rm chemi}$ from the position of the new stellar particle
 and then  count
the number of the neighbor gas particles, $N_{\rm gas}$.
This  $R_{\rm chemi}$ is referred to as 
chemical mixing length in the present paper,
and represents the region within which the neighbor
gas particles are polluted by metals ejected from the new stellar particle.
The value of  $R_{\rm chemi}$  relative to the typical size of
a galaxy is set to be 0.01. 
Next we assign the metallicity of original
gas particle to  the new stellar particle and increase 
the metals of the each neighbor gas particle according to the following 
equation about the chemical enrichment:
  \begin{equation}
  \Delta M_{\rm Z} = \{ Z_{i}R_{\rm met}m_{\rm s}+(1.0-R_{\rm met})
 (1.0-Z_{i})m_{\rm s}y_{\rm met} \}/N_{\rm gas} 
  \end{equation}
where the $\Delta M_{\rm Z}$ represents the increase of metal for each
gas particle. $ Z_{i}$, $R_{\rm met}$, $m_{\rm s}$,
and $y_{\rm met}$  in the above equation represent
the metallicity of the new stellar particle (or that of original gas particle),
the fraction of gas returned to interstellar medium,  the
mass of the new star, and the chemical yield, respectively.
The value of the $R_{\rm met}$ and that of $y_{\rm met}$ are set to
be 0.2 and 0.03, respectively.
Furthermore, the time, $t_{i}$, 
when the new stellar particle is created, is 
 assigned to the new stellar particle in order to calculate the 
photometric evolution of merger remnants, as is described later.
To verify the accuracy of the above  treatment 
(including numerical code) for 
chemical enrichment process,
we checked whether or not the following conservation law of chemical 
enrichment is satisfied for each time step in each test simulation:
  \begin{equation}
 \sum_{\rm star} m_{\rm s}Z_{i} + \sum_{\rm gas}
  m_{\rm g}Z_{i} = y_{\rm met}
 \sum_{\rm star} m_{\rm s}
  \end{equation}
where  $ m_{\rm g}$,  $ m_{\rm s}$, and $ Z_{i}$
 are  the mass of each gas particle,
that of each stellar one, and the metallicity
of each particle,  respectively, and the  summation ($\sum$) is done
for all the gas particles or stellar ones. 
Strictly speaking, the above equation  holds when the value of
$R_{\rm met}$ is 0.0.
Thus, in testing the validity of the present code of chemical enrichment,
we set the  value of  $R_{\rm met}$ to be 0.0 and then perform a simulation
for the test.
We confirmed that the above equation is nearly exactly satisfied
in our test simulations and furthermore that even if the  $R_{\rm met}$ is
not  0.0, the difference in the value of
total metallicity between the left and right side in the above equation
is negligibly small.
Our numerical results do not depend so strongly on
$R_{\rm chemi}$, $R_{\rm met}$, $y_{\rm met}$ within
plausible and realistic ranges of these parameters.

\subsubsection{Population synthesis}

It is assumed in the present study
that the spectral energy distribution (SED) of a model galaxy is 
a sum of the SED of  stellar particles. 
The SED of each  stellar particle is assumed to be  
a simple stellar population (SSP) that is  
a coeval and chemically homogeneous assembly of stars. 
Thus the monochromatic flux of a galaxy with  age $T$,
$F_{\lambda}(T)$,  is described as 
\begin{equation}
F_{\lambda}(T) = \sum_{\rm star} F_{{\rm SSP},\lambda}(Z_{i},
{\tau}_{i}) \times m_{\rm s} \; ,
\end{equation}
where $F_{{\rm SSP},\lambda}(Z_{i},{\tau}_{i})$ and $m_{\rm s}$ 
 are  a monochromatic flux of SSP 
of age ${\tau}_{i}$ and metallicity $Z_{i}$, where suffix $i$ identifies 
each stellar particle,  and 
mass of each stellar  particle,  respectively.
The age of SSP, ${\tau}_{i}$, is defined as ${\tau}_{i} = T - t_{i}$, 
where $t_{i}$ is the time when a gas particle is
converted  into a stellar one.
The metallicity of SSP is exactly the same
as that  of the stellar particle, $Z_{i}$, and the summation ($\sum$) in
equation (4) is done  for all 
stellar particles in a model galaxy.

A stellar particle is assumed to be composed of stars whose
age and metallicity are exactly the same as those of the stellar particle
and 
the total mass of the stars is set to be the same as that of
the  stellar particle.
Thus the monochromatic flux of SSP at a given wavelength is defined as
\begin{equation}
F_{{\rm SSP}, \lambda}(Z_{i},{\tau}_{i}) = \int_{M_L}^{M_U} 
\phi (M) f_{\lambda}(M, {\tau}_{i}, Z_{i}) dM \; ,
\end{equation}
where $M$ is mass of a star, $f_{\lambda}(M, {\tau}_{i}, Z_{i})$
 is a monochromatic flux 
of a star with mass $M$, metallicity $Z_{i}$ and age ${\tau}_{i}$.
$\phi (M)$ is a initial mass function (IMF) of stars and 
$M_U$, $M_L$ are upper and lower mass limit of IMF, respectively. 
We here adopt the Salpter IMF with $M_U$ = 120$M_{\odot}$ and
$M_L$ = 0.1 $M_{\odot}$. 
In this paper, we use the $F_{{\rm SSP}, \lambda}(Z_{i}, {\tau}_{i})$ 
of GISSEL96, which is the latest version of Bruzual \& Charlot (1993).

\placefigure{fig-1}

\subsection{Model of internal dust extinction}

\subsubsection{Derivation of a SED modified by dust effects}

Using the derived each stellar particle's SED not corrected by dust 
($ F_{{\rm SSP},\lambda}(Z_{i}{\tau}_{i}) \times m_{\rm s}$
shown in equation (4) ) and
stellar and gaseous distribution, 
we can obtain the  SED of a galaxy corrected by dust.
From now on the SED of a stellar particle, $ F_{{\rm SSP},\lambda}(Z_{i},
{\tau}_{i}) \times m_{\rm s}$,
 is referred to as 
$F_{sed,i}$ for convenience.
We first calculate dust extinction of star light for $each$  stellar particle
and dust temperature for $each$ gaseous particle, based on the three dimensional
spatial distribution of stellar and gaseous particles.
We  then  sum each stellar particle's  SED corrected by dust extinction and the
dust reemission of each gaseous particle.
The method to derive the dust extinction and reemission for each particle 
consists of the following three steps.
Firstly we assume that each stellar particle with the SED
$F_{sed,i}$ emits  rays  with   the SED for each of the rays
equal with  $F_{sed,i}$/$N_{ray,i}$,
where $N_{ray,i}$ is the total number of the rays.
We generate random numbers for each ray and thereby determine the direction
of each ray from the stellar particle. 
Figure 1 represents a schematic explanation of this 
first step.
$N_{ray,i}$ is set to be 10 
for all stellar particles in the present study.
Secondly, we investigate  whether each of rays emitted from a
stellar particle can penetrate a gas  particle for  all gas 
particles and thereby estimate dust extinction of 
stellar light for each of the rays.
This investigation and estimation is done for all stellar particles.
In this second step, we regard a ray as penetrating a gas particle
and thus being affected by internal dust extinction if
the ray passes through the region within the gas cloud radius ($r_{cl}$). 
This second step is also schematically shown in Figure 1.
Here gas clouds with the masses of  $m_{cl}$
are assumed to be spherical and the cloud
radius $r_{cl}$ is changed following the relation (Kwan \& Valdes 1987),

\begin{equation}
 r_{cl}= k_{cl} \times 47 {(\frac{m_{cl}}{4 \times 10^{6} \rm M_{\odot}})}^{1/3}
  pc \; 
\end{equation}
Although physical processes associated with the growth and the disruption
of gas clouds have been investigated (e.g., Kwan 1979;
Kwan \& Valdes 1987; Olson\& Kwan 1990a, b), it is not so clear
which the most probable and physically reasonable value of $k_{cl}$ is for
evolving galaxies.
A probable value of $k_{cl}$ is  $\sim$ 1.0 for  disk galaxies
(Kwan \& Valdes 1987) whereas the
typical cloud mass and size  depend strongly on time in merging
galaxies owing to coalescence and disruption of 
gas clouds (Olson\& Kwan 1990a, b). 
Olson\& Kwan (1990b) demonstrated that although
most of the cloud-cloud collisions induced by galaxy merging
is disruptive,  
several very large gas clouds with masses greater than $10^7$
$M_{\odot}$ can form after merging owing to successive coalescence
of colliding clouds. 
Furthermore the density  of gas clouds  in ULIRGs  considered to
be ongoing mergers is suggested to be  very high from  
that in disk galaxies (Aalto et al. 1991a, b),
which implies that the size and the mass might be also
different from those of isolated disk galaxies.
Accordingly it is reasonable for us to  assume that $k_{cl}$ is
a free parameter for the present study.
In the present paper, however, we show only the results for the model
with $k_{cl}$ = 4.68.
The $k_{cl}$  dependence in major mergers will be
described in our future papers. 

Absorption of stellar  light for each of rays from
a stellar particle   is modeled according
to the following reddening formulation (Black  1987; Mazzei et al. 1992); 
$E(B-V)=N(\rm H)/4.77 \times {10}^{21} {\rm cm}^{-2} \times ({\it Z}_{\rm g}/0.02)$,
where $N(\rm H)$ and ${Z}_{\rm g}$ are 
gaseous column density and gaseous metallicity, respectively.
Since dust absorption is estimated for each of gas particles
penetrated by a ray emitted from a stellar particle,
the above $N(\rm H)$ and ${Z}_{\rm g}$  correspond to
column density of each gas cloud particle (${N(\rm H)}_{i}$) and metallicity of
the particle ($Z_{\ i}$), respectively. 
The $Z_{\ i}$ is derived from our numerical simulations
and the ${N(\rm H)}_{i}$ is derived from the  relation
between cloud mass   
and size shown in equation (6).
Using the derived redding $E(B-V)$ and extinction law by Cardelli et al. (1989),
we adopt the so-called screen model for rays
from a stellar particle and  calculate the dust
absorption of the stellar particle.
Thirdly,
by  assuming  the modified black body radiation 
with  the emissivity ($\epsilon$) law 
$\epsilon \propto {\nu}^{{\alpha}_{\rm em}}$, 
we  determine the dust temperature of a gas particle such  that
total energy flux of dust absorption is equal to that of dust reemission. 
In the present study, we  show the results of
a model with ${\alpha}_{\rm em}$ = 2.
We here do not include the albedo  for dust grains, which means that only stars and new stars
heat dust.
Thus the SED of a galaxy consists of the stellar continuum 
modified by dust extinction and the reemission of dusty gas.

The method  adopted in the present numerical code 
calculating dust absorption and reemission
is essentially  the same as the so-called ray tracing method
 adopted in previous theoretical calculations on radiative
transfer problems 
(e.g. Witt 1977; Efstathiou \& Rowan-Robinson 1990).
This paper is the  first step toward 
a sophisticated combination of N-body simulations with the already
existing ray tracing method.
Therefore there are still some problems of the present new code in
calculating correctly the SED within  feasible time scale
for a simulation with the total particle number approximately
equal to $10^{5}$.
The most significant problem is on the time which we spent in calculating
the SED. The time necessary for the SED calculation for each time step
in a simulation can be roughly scaled to $n_{s} \times n_{g} \times
n_{ray}$, where $n_{s}$, $ n_{g}$, and  $n_{ray}$  
are
total particle number of stars, that of gas, and that of rays
emitted from each stellar particle.
Therefore it is very time-consuming to derive a SED
in each time step of a simulation, in particular,  for large N-body simulations
with both  $n_{s}$ and $n_{g}$ larger than $10^{5}$:
It is obviously our future work to develop a new method for 
calculating SEDs of such large N-body systems.
Furthermore, our SED calculation becomes also time-consuming,
unless we carefully choose the parameter value of  $n_{ray}$.
Although we can more correctly estimate SEDs of galaxy mergers
with dusty starburst for the models with larger $n_{ray}$,
we should adopt a feasible and reasonable value of $n_{ray}$,
considering that the time for our SED calculation is roughly
proportional to $n_{s} \times n_{g} \times n_{ray}$.
In order to determine a plausible value of $n_{ray}$,
we investigated the dependence of a SED on $n_{ray}$ (1 $\le$ $n_{ray}$
$\le$ 50) in a dusty starburst merger
model and found that if $n_{ray}$ $\ge$ 10, the differences in SEDs 
between models with different $n_{ray}$ become negligibly small. 
We thus adopt $n_{ray}$ = 10 as a feasible and plausible value
for our SED calculations in the present study.
Although our new numerical method allows  us to derive a SED
for an arbitrary distribution of stellar and gaseous component
for a dusty galaxy in numerical simulations,
it still has some problems such as those described above.
Both our method and previous ones 
(i.e., those
by which numerical studies can precisely solve radiative transfer processes
and estimate galactic SEDs 
within a certain spherical symmetric approximation or axisymmetric one) 
have advantages and disadvantages, 
and therefore we stress that the present numerical study is complementary
to previous ones.

\subsubsection{Limitations of the present code}

Our numerical  model can derive a SED for a galaxy with an arbitrary 
distribution of stellar and gaseous components
and accordingly enable us to examine a  SED
of a galaxy with very peculiar mass distribution, such as galaxy mergers
and forming galaxies.
However, our model has the following four limitations in investigating
the time evolution of SEDs in galaxies.
Firstly, we do not include physical processes related to
dust production, grain destruction, and grain growth in interstellar 
medium in the present model
and thus cannot follow the time evolution of dust composition.
Dwek (1999) developed a new self-consistent model for the evolution
of the compositions and abundances of elements and the dust 
in galaxies by including condensation, accretion, and destruction
processes of grains in the model.
By using this model, 
Dwek (1999) investigated the time evolution of dust composition
(e.g., whether interstellar gas is silicate rich or not)
and suggested that extinction law, which is an important 
factor for calculating a galactic SED,
depends on the evolution. 
The method developed by Dwek (1999) is mainly for one-zone
models of galactic chemical evolution, accordingly it
can not be so simply applied to  N-body numerical simulations.
It is thus our future study to combine the method provided by  
Dwek (1999) with our numerical code and thereby to
derive a galactic SED in a more self-consistent manner. 
Furthermore Chang, Schiano, \& Wolfe (1987) 
investigated how the radiation from QSOs can affect 
the hydrodynamical evolution of dusty interstellar gas,
considering ion-field emission, ion sputtering,
coupling between gas ions and dust grains, photoionization,
and photoelectrical processes.
They found that owing to the high sputtering on the grain surface
in dusty interstellar gas, most of the grains inside 7 kpc of
a QSO host galaxy can be destroyed within $\sim$ 3 $\times$ 
$10^7$ yr.
Although their results are not directly related to
the dust destruction processes in dusty starburst galaxies,
we consider that the essence of their results can be applied to the case of
starburst galaxies.
The time scale for intense starburst in ULIRGs formed by
major mergers  is suggested to be $\sim$ 
$10^8$ yr and the total luminosity 
of ULIRGs are similar to QSOs (e.g., Sanders et al. 1988),
and accordingly the radiation from nuclear strong starburst can
also destruct the dust grains in ULIRGs.
If the dust grains are efficiently destructed by radiation of starburst,
the total flux of far-infrared reemission from dusty starburst
can be greatly decreased.
Our present model does not include this effects of starburst radiation
on dust grains, we stress that the total value  of infrared and far-infrared
luminosity can be overestimated in major mergers with dusty starburst.

Secondly,  we do not consider the effects of small grains
with the size between 1 and 10 nm, and consequently
can not precisely estimate a SED around $10^{5}$
$ \rm \AA $
in a dusty system.
Devriendt, Guiderdoni, \& Sadat (1999) included
contributions from polycyclic aromatic hydrocarbons (PAHs),
small grains, and big ones in their model and succeeded
in reproducing reasonably well the SEDs of several ULIRGs.
To construct multi-component dust model is accordingly
the next step of our studies toward the more precise
estimation of SEDs of dusty galaxies.
Thirdly, the parameter $k_{cl}$ is fixed at a possibly
reasonable value during a simulation,
accordingly our model cannot follow the time evolution
of cloud size and mass.
Witt \& Gordon (1996) demonstrated that the change
of the structure of interstellar medium 
in a dusty system  provides
the change in effective optical depth and consequently
controls  the SED of the system.
They furthermore suggested that 
a breakup of large interstellar cloud complexes
into numerous smaller clouds and a intercloud medium
of enhanced density in a galaxy merger
can greatly affect the SED of the merger.
Fourthly, we do not include the effects of albedo at all
in the present study, though several authors
have already investigated in detail the effects of dust albedo
(e.g.,  multiple scattering of stellar light)  on
SEDs in dusty systems (e.g., Witt \& Gordon 1996; Gordon et al. 1997).
Although the total number of parameters increase if we
include the effects of dust albedo in our numerical simulations,
it is our future work to investigate the importance
of dust albedo in determining galactic SEDs. 
Thus our numerical results should be carefully interpreted owing to the above
four limitations of the code used in the present study.

\subsection{Main points of analysis}

Mainly important advantages of the present study
are the following two.
Firstly  we can investigate simultaneously morphological, structural,
kinematical, and photometric  properties
at a given time step in a galaxy merger with dusty starburst
based on the derived SED of the merger.
Secondly, by using the SED derived
for a merger at a given z  and considering the effects of k-correction,
we can investigate how the morphological and photometric
properties of the merger change with redshift. 
Accordingly we mainly  investigate the time evolution 
of morphology, star formation, global colors, luminosity
and $A_{\rm V}$ in a dusty starburst galaxy merger.
Furthermore we investigate two-dimensional distributions
of colors, luminosity, and $A_{\rm V}$,
which have not been investigated at all in
previous theoretical studies of dusty galaxies.  
In order to discuss apparent morphology of intermediate
and high $z$ dusty galaxies, 
 we also investigate how dust-enshrouded starburst galaxy mergers
 at z = 0.4, 1.0, and 1.5 can be seen in the Hubble Space
Telescope ($HST$). 
 The method to construct the  synthesized $HST$ images of galactic morphology
 in the present study is basically the same as that described by Mihos (1995)
 and Bekki, Shioya, \& Tanaka (1999).
 We adopt  Madau's model (Madau 1995) for intervening absorption of 
neutral hydrogen and consider k-correction of galactic SEDs
in order to derive the synthesized  images.
For comparison, we also show how 
dust-enshrouded starburst galaxy 
mergers at  z = 0.4, 1.0, and 1.5 can be seen 
in the SUBARU which is a Japanese large (8.2m) grand-based telescope.
The results on the  synthesized $HST$ and SUBARU images are 
presented in \S 3.3 and 
for discussing 
the origin of high $z$ faint SCUBA sources and EROs. 
In order to calculate the SEDs of the merger
model at each redshift,
we assume that mean ages of old stellar components
initially in  a merger progenitor   disk at the redshift
z=0.4, 1.0, and 1.5 are 7.14, 3.80, and 2.46 Gyr, respectively.

We here describe mainly the results of only one merger
model with $k_{cl}$ = 4.68 (i.e., $r_{cl}$ is 0.011 in our units
corresponding to 200 pc) 
because the main purpose of the present study is not
to give dependences of SEDs on physical parameters of galaxy mergers
but to demonstrate usefulness of the present code in studying theoretically
dusty starburst galaxies.
From now on this mode is  referred to as the standard model.
We do not intend to change the values of important free parameters
such as $M_{\rm d}$,     $M_{\rm t}$,  $r_{\rm in}$, $r_{\rm p}$,
$\phi_{i}$,  $\theta_{i}$, and $f_{g}$ in the present study.
Dependences of the evolution of
dusty starburst mergers on parameters will be described
in our future papers (Bekki \& Shioya 2000b). 
Although each of the results described in the following sections
have much implications on the nature of ULIRGs
and can be  compared with up-to-date observational results
such as near-infrared colors (e.g., Scoville et al. 1999),
surface brightness distribution (e.g., Sanders et al. 1999;
Scoville et al. 1999), and high resolution optical/near-infrared
images of ULIRGs (e.g., Surace, Sanders, \& Evans 1999),
we do not intend to discuss so extensively the origin and the nature
of ULIRGs in the present study:
This is simply because the most important purpose
of the paper is to demonstrate the importance and the usefulness
of the present code in investigating dusty starburst galaxies.
We will discuss the origin of ULIRGs in our future  papers.
 In the followings, the cosmological parameters
 $\rm H_0$ and $\rm q_0$ are  set to be 50 km s$^{-1}$ Mpc$^{-1}$ and 0.5
 respectively.

\placefigure{fig-2}
\placefigure{fig-3}

\section{Result}

\subsection{Evolution of morphological properties and SEDs}

 Figure 2 and  3 describe time evolution of morphology of the standard 
model 
for each of four components, dark halo,  stars initially located within
two disks, gas, and new stars formed during galaxy merging.
From now on, for convenience, the time $T_{m}$ represents the time that has
elapsed since the two disks begin to merge. 
As galaxy merging proceeds,  the two disks are strongly disturbed 
to form a long tidal tail in the disk orbiting in a prograde sense
at 0.6 $<$ $T_{m}$ $<$ 1.1  Gyr.
A remarkable
tidal tail is not developed in the disk orbiting in a retrograde sense.
This  one long tidal arm is characteristics of prograde-retrograde
mergers.
The two disks finally sink into the center of massive dark halos
owing to dynamical friction during merging and consequently
are completely destroyed by violent relaxation of 
galaxy merging (1.1 $<$ $T_{m}$ $<$ 1.7  Gyr).
As a result of this, the two disks form an elliptical galaxies
with the structure and kinematics similar to the observed 
ones ($T_{m}$ = 1.7 Gyr).
During violent major merging, interstellar gas is very
efficiently transferred to the central region of the two disks 
owing to gaseous dissipation of colliding gas clouds
and gravitational torque. 
Gas accumulated in the central region of the merger is
then consumed by massive starburst and consequently
converted into new stars. As is shown in Figure 4,
the star formation rate becomes maximum 
($\sim378$ $M_{\odot}$ $\rm yr^{-1}$)
at $T_{m}$ = 1.3 Gyr when two disks of the merger become
very close to suffering from violent relaxation.
The maximum star formation rate is roughly two orders of
magnitude larger than the mean star formation rate
of an isolated disk in the present study and comparable
to that required for explaining the strong infrared luminosity
observed in ULIRGs. 
After the strong starburst, star formation  rate rapidly declines
($\sim$ 1 $M_{\odot} {\rm yr}^{-1}$)  
within less than 1 Gyr after $T_{m}$ = 1.3,
essentially because most of the gas is consumed up by
the starburst.
These results are qualitatively consistent with those in Mihos \&
Hernquist (1996).

We here stress that the above one-time starburst in the late phase of
major merging  is applied only to the present prograde-retrograde
merger with the adopted inclination of two disks and internal
structure of the disks. 
Mihos \& Bothun (1998) demonstrated that physical details
such as internal structure of merger progenitor disks and
their initial gas mass fraction are important
for the star formation history and the luminosity evolution
in ULIRGs, by using imaging Fabry-P$\acute{\rm e}$rot observations on
four ULIRGs.
They furthermore found  that the spatial
distribution of H$\alpha$ emission from
starburst regions is very diverse in the four ULIRGs 
(i.e., some ULIRGs show the strong H$\alpha$ emission
only in the central region
and H$\alpha$ emission is quite extended in some ULIRGs) 
and accordingly suggested that several different factors 
play a role in triggering starburst in ULIRGs.
Considering these important observational results,
we suggest that the results on the star formation history
and the resultant luminosity evolution described
in the present study are  only true for some ULIRGs.

Figure  5 and 6
 give mass distribution projected onto $x$-$y$ plane (orbital plane)
and $x$-$z$ plane
at $T_{m}$ = 1.1 (pre-starburst phase) and 1.7 Gyr (post-starburst phase)
for stars, gas, and new stars.
New stars formed mainly by secondary massive starburst
are more  compactly distributed in the merger
than old stars initially located within disks
at the pre-starburst (weak starburst) epoch 
when the two cores in the merger have not yet merged with each other
to form an elliptical galaxy.
This is  essentially
because new stars experience much more gaseous dissipation
when they were previously gaseous components.
Consequently new stars with younger ages are
more heavily obscured by dusty gas than  old stars  during galaxy merging.
This result that young stellar components can be  preferentially
obscured by dust during evolution
of galaxy mergers is suggested to be very important for understanding
the nature of post-starburst galaxies detected by Smail et al. (1999)
in SCUBA surveys of intermediate redshift clusters of galaxies
(Shioya \& Bekki 2000).
Three very small compact stellar clusters composed mainly of
new stars (and gas) can be seen above the two cores at  
$T_{m}$ = 1.1.  
This results suggests that efficient star formation 
during  galaxy merging
can occur not only in the central part of a merger
but also in stellar clusters located in 
the outer part of the merger.
As is shown in Figure 6, the morphology of the merger
only $\sim$ 0.4 Gyr  after the maximum starburst
looks like an elliptical galaxy,
which implies that the time scale within which 
a merger can be seen as an ULIRG with the very peculiar morphology
is very short (less than 0.5 Gyr).

Figure 7 describes the SEDs of the merger which we derive
based on the mass distribution of stellar and gaseous component
shown in Figure 2 and 3 and the age and metallicity distribution
of stellar populations of the merger
at $T_{m}$ = 0.6, 1.1, 1.3 (the epoch of maximum starburst), 
1.7, 2.3, and 2.8 Gyr.
We can clearly observe how the dust extinction
of interstellar gas can change the shape of the SED
of the merger at each time by comparing the results 
of the model with dust extinction with those  without dust extinction.
The UV flux with the wavelength
less than 3000 $\rm \AA$ rapidly increases
during 1.1 $<$ $T_{m}$ $<$ 1.3 Gyr owing to a large number 
of young massive stars formed by the strong starburst 
whereas it decreases during 1.3 $<$ $T_{m}$ $<$ 2.8 Gyr
because of very small star formation rate after the starburst 
in the model without dust extinction.
The infrared and
submillimeter fluxes become larger during 1.1 $<$ $T_{m}$ $<$ 1.3 Gyr
in the model with dust extinction.
This is firstly because star formation
rate, which is closely associated
with the total mount of stellar light absorbed by
interstellar dust,
becomes considerable  higher owing to the efficient gas transfer to
the central region of the merger and secondly because
the  density of dusty gas becomes also
very high in the later phase of the merging
(1.1 $<$ $T_{m}$ $<$ 1.3 Gyr)
so that the gas can heavily obscure the strong starburst.
After the maximum starburst at $T_{m}$ = 1.3 Gyr,
both infrared  and submillimeter fluxes rapidly decline 
(1.3 $<$ $T_{m}$ $<$ 1.7 Gyr).
This is principally because most of interstellar gas indispensable for
strong starburst and dust obscuration is rapidly 
consumed by star formation in the  merger till $T_{m}$ $\sim$ 1.7 Gyr. 
Thus the time evolution of SED of a merger depends strongly on
that of the star formation rate, which is basically controlled
by dynamical evolution of the merger. 

\placefigure{fig-4}
\placefigure{fig-5}
\placefigure{fig-6}
\placefigure{fig-7}

Figure 8 and 9 describe the mass distribution and
the SED 
at the epoch of maximum starburst in the standard model,  respectively. 
Figure 8 clearly
shows that both new stars and gas are more centrally concentrated 
than old stellar components,
which means that star light from new stars formed by the
massive starburst are more heavily obscured  by dusty interstellar medium
than that of old stellar components.
The SED at $T_{m}$  = 1.3 Gyr in Figure 9 
is rather  similar to the observed SED of typical ULIRGs,
which implies that the present dusty starburst merger model can be 
observed as an ULIRG in the late of galaxy merging.
These  results in Figure 8 and 9
clearly demonstrate that starburst population in the merger
is so heavily obscured by dusty interstellar medium  (the
mean $A_{V} \sim 2.46$ mag) 
that  the dust reemission
in far-infrared  ranges ($L_{IR}$) 
becomes very strong at the maximum starburst
($L_{IR}=1.59 \times 10^{12} \rm L_{\odot}$). 
The reasons  for this heavy dust extinction are  firstly that 
column density of dusty interstellar medium  becomes
extremely high owing to the strong central accumulation of gas 
and secondly  that the central gaseous metallicity, which is a measure
of total amount of dust,
also becomes rather large ($\sim 0.04$)  because of efficient and rapid chemical evolution
in galaxy merging with secondary starburst.

Based on the derived mass distribution and SEDs at
the maximum starburst (at ULIRG phase),
we can examine two-dimensional distribution of
$K$ band surface brightness (Figure 10), rest-frame $R-K$ color (11), 
and $A_{V}$ (12), all of which have been
observationally investigated in a very extensive manner 
for the better understanding of the nature
of ULIRGs.
As is shown in  Figure 10,
near-infrared surface brightness 
is the highest (the smallest in number)  in the central region within which  
most of young bright massive stars are located.
This is essentially because most of young stars  
are formed  in the central region at $T_{m}$ = 1.3 Gyr.
The $R-K$ color is also the largest in the central
part of the merger (Figure 11),
because the central starburst components
are most heavily obscured by dust owing to
the very high gaseous  density there.
Accordingly, this result implies that
an ULIRG formed by major merging can show  negative color
gradients. 
As is shown in Figure 12, 
$A_{V}$ is larger in
the central region of the merger
than in the outer part, 
which 
indicates that dust absorption
and its reemission is larger in the central region.
This result is consistent with those described in Figure 11.
 
\placefigure{fig-8}
\placefigure{fig-9}
\placefigure{fig-10}
\placefigure{fig-11}
\placefigure{fig-12}

\subsection{Photometric and color evolution}
Figure 13 describes the time evolution
of absolute magnitude from optical to near-infrared wavelength
($M_{V}$, $M_{R}$, and $M_{K}$)
for the standard  model with and without dust extinction.
As galaxy merging proceeds,
$M_{V}$, $M_{R}$, and $M_{K}$ gradually rise up  owing
to the increase of star formation rate (1.1 $<$ $T_{m}$ $<$ 1.3 Gyr).
These $M_{V}$, $M_{R}$, and $M_{K}$ then become
$-21.0$, $-21.5$, and $-24.2$ mag, respectively,
at the epoch of maximum star formation rate ($T_{m}$ = 1.3 Gyr)
in the model with dust extinction.
The difference in magnitude between the models with and without
dust extinction at the epoch
of maximum starburst 
is the largest for $V$ band (2.5 for $V$, 2.1 for $R$
and 1.3 for $K$ band),  which reflects the fact that stellar light
with shorter wavelength can be more greatly  absorbed   by dust.  
The mean value
of $M_{K}$ in ULIRGs ($-25.2$ mag; Sanders \& Mirabel 1996)
is about 1 mag brighter (smaller in number)  than the 
$M_{K}$ of the present merger at $T_{m}$ = 1.3 Gyr,
which probably means that initial total mass
of the host disk of a typical ULIRG  is about 2 times  
larger than that of the Galaxy
adopted as a merger progenitor disk in the present study.
Furthermore, irrespectively of wavelength,
the difference in absolute magnitude between the two models
in the preset study 
is the largest at the epoch of maximum starburst. 
This is essentially because young and very luminous stellar
components formed by starburst are the most heavily
obscured by dust at the maximum starburst epoch 
when a larger amount of dusty gas is transfered
to form very high density region around the compact starburst. 
As the star formation rate rapidly declines,
$M_{V}$, $M_{R}$, and $M_{K}$ also rapidly decline (1.3 $<$ $T_{m}$ 
$<$ 2.8 Gyr) because of the death of young and luminous stellar components
(OB stars formed in starburst) and the aging of stellar populations.
The present results thus  suggest that time variation 
of absolute magnitude during galaxy merging becomes moderate  
owing to dust extinction.

Figure 14 shows the time evolution of $V-I$, $I-K$, and $R-K$
colors in the standard model  with and without  dust extinction.
Although the colors 
in the model without extinction  become very blue
at the epoch
of maximum starburst,
these colors are changed into redder ones for a starburst galaxy
because of the heavy dust extinction in the merger. 
The color difference between the models with and without dust extinction
is the most remarkable 
at the epoch of maximum starburst. 
Figure 15  describes the time evolution 
of infrared flux at 60 and 100 $\mu$m 
and submillimeter one at 450 and 850 $\mu$m  
for the model with  dust extinction.
Clearly  these fluxes resulting mainly from dust
reemission become maximum 
when the magnitude of starburst
is the largest and the  density of dusty gas in the central region
of the merger is also considerably high.
Furthermore these fluxes become an order of
magnitude smaller than the maximum values  within $\sim$ 1  Gyr
after the maximum starburst.
This result implies that the time scale during which a dusty starburst
merger can be identified as  a submillimeter source detected
recently by the SCUBA is very short (an order of $10^{8}$ year). 
As is shown in Figure 16, the dust temperature of the model
becomes maximum ($\sim$ 32 K) when the star formation rate becomes
maximum and  total amount of radiation from young massive stars  
formed by starburst becomes the largest. 
We here note that the dust temperature 
is the mean value for all gas particle for each of which the dust temperature
is calculated.
Figure 16 implies that the dust temperature in mergers with dusty starburst 
depends strongly on the strength of starburst.

\placefigure{fig-13}
\placefigure{fig-14}
\placefigure{fig-15}
\placefigure{fig-16}

\subsection{Evolution with redshift}

Figure 17, 18, and 19 describe how the difference in the  epoch
of galaxy merging affects the global colors 
($V-I$, $I-K$, and $R-K$, respectively) in observed frame 
for the standard model. For clarity, we give the dependence of colors
on the merging epoch z only for the three different phases
of the merger, the pre-starburst epoch ($T_{m}$ = 1.1 Gyr),
the maximum starburst one (1.3),
and the post-starburst one (1.7). 
As is shown in Figure 17, optical color $V-I$ in observed frame
is the reddest at the epoch of post-starburst ($T_{m}$ = 1.7 Gyr)
at each redshift (i.e., at the epoch of galaxy merging). Furthermore, 
the $V-I$ color difference between different redshifts
is about 0.7 mag for the epoch of maximum starburst
when infrared flux becomes the largest.
As has been shown in  Bekki et al (1999),  
the submillimeter flux at 850 $\mu$m at
z $>$ 1.0 for
the maximum starburst phase
($T_{m}$ = 1.3 Gyr) 
is  a few mJy that is well above the current detection limit
of the SCUBA ($\sim$ 2 mJy).
These  results thus imply that the observed color difference in $V-I$
for the faint SCUBA sources (Smail et al. 1998) is due partly
to the difference in redshifts between dusty starburst galaxies
detected by the SCUBA.
Irrespectively of redshifts, near-infrared colors
$I-K$ and $R-K$ are the reddest for the post-starburst phase
(Figure 18 and 19). 
Furthermore these colors are redder in higher z
for the post-starburst phase ($T_{m}$ = 1.7 Gyr) because of the k-correction
of stellar populations in this phase. 
It should be noted here that the $R-K$ color for the  post-starburst phase 
becomes larger than $\sim$ 6.0 around z = 1.5. 
These results imply that an ERO with the $R-K$ color larger than 6.0
is more likely to be formed in the post-starburst phase of a
gas-rich major merger with the strong dusty starburst and
at  higher redshifts (z $>$ 1 $\sim$ 2).  
These results on the redshift evolution of global colors 
are only for one merger model.
Therefore, we stress that although the above results are useful
and helpful for understanding  the nature 
of the SCUBA sources and EROs, the results
can be true only for some high z dusty mergers.
It is our future work to investigate a much larger number of dusty
merger models and thereby clarify the nature and the origin
of high z dusty starburst galaxies such as the faint SCUBA sources
and EROs.

\placefigure{fig-17}
\placefigure{fig-18}
\placefigure{fig-19}

\section{Discussion}

One of advantages of the present new numerical code is 
that we can investigate both dynamical and photometric evolution
of dusty starburst galaxies 
in an explicitly self-consistent manner.
Therefore we can address some important questions that
classical one-zone models on galaxy evolution
have not yet answered so clearly for the origin of dusty starburst
galaxies.
Here, we  particularly enumerate  three problems  which 
the newly developed code is rather helpful and 
useful for clarifying
and thus will be discussed in our future papers.

\subsection{Origin of faint SCUBA sources}

 Recent observational studies with 
 the SCUBA 
have revealed
 possible candidates of heavily dust-enshrouded starburst galaxies at intermediate
 and high redshift, which could be counterparts of low redshift  
 ULIRGs (Smail, Ivison, \& Blain 1997;
 Barger et al. 1998;
Hughes et al. 1998; Smail et al. 1998; Ivison et al. 1999; Lilly et al.  1999).
 Although optical morphology of these submillimeter extra-galactic sources should be treated with
 caution owing to the absence of high resolution sub-mm imaging capability (Richards 1999), 
 more than 50 \% of those are  suggested  to 
 show the indication of galactic interaction
 and merging (Smail et al. 1998).
 Furthermore observational studies of an extremely red object ERO J 164502-4626.4 (HR 10)
 with the redshift of 1.44 by the Hubble Space Telescope and the SCUBA have
 found that this high redshift galaxy  is also a dust-enshrouded
 starburst galaxy with clear indication of galaxy merging/interaction 
 (Graham \& Dey 1996; Cimatti et al. 1998; Dey et al. 1998).
 Although
 low redshift ULIRGs are generally considered to be ongoing
 galaxy mergers with the triggered  prominent nuclear activities (starburst or AGN)
 heavily obscured  by dust (Sanders et al. 1988; Sanders \& Mirabel 1996),
the origin of the faint SCUBA sources is not so clearly understood. 
In particular, it is not clear whether
the observed very large  submillimeter luminosity,  
which could result either from active galactic nuclei (AGN)
obscured by dust or from dusty starburst,
is due essentially to physical processes associated closely with 
major galaxy
merging.

Concerning this problem, 
one of tests to assess the validity of merger scenario
of the faint SCUBA source formation is to compare
the observed morphological, structural, and kinematical
properties of the sources with
those of a certain theoretical merger model
$at$ $the$ $considerably$ $strong$ $starburst$ $epoch$ 
when submillimeter flux of the merger can exceed the
detection limit of the SCUBA.
Therefore, theoretical studies should investigate
simultaneously the time evolution of 
submillimeter flux at 850 $\mu$m, morphology,
structure, and kinematics of a dusty major galaxy merger
in order to compare the results with the corresponding observational
ones.
Although classical one-zone models
can investigate photometric evolution of dusty starburst
galaxies at each wavelength and thus have contributed
greatly to the understanding of dusty galaxies (e.g., Mazzei et al 1992),
 they cannot predict dynamical evolution of galaxies simultaneously.
The present model, on the other hand,
can predict not only photometric evolution from UV 
to submillimeter wavelength but also  dynamical evolution 
in a dusty galaxy.
For example, as is shown in the present study,
we can predict optical and near-infrared morphology of a merger
with the  850 $\mu$m flux larger  than 2 mJy (the current detection
limit of the SCUBA) at a given redshift. 
Thus we expect that future theoretical
studies with our  new code will provide
valuable clues to the origin of the faint SCUBA sources.

\subsection{Origin of EROs}

 Recent observational studies have discovered
a significant number of 
high z EROs with $R-K$ $\ge$ $5 \sim 6$
(Elston et al.
1988; Cimatti et al. 1998;
Dey et al. 1999; Ben\'\i tez et al. 1998; Cimatti et al. 1999;
Smail et al. 1999; Soifer et al. 1999),
and accordingly the nature and the origin of EROs 
have now been discussed very extensively both in observational
studies and in theoretical ones.
Thompson et al. (1999) argued
that  EROs
are  most likely to lie in the redshift range
1 $<$ $z$ $<$ 2 and they represent an important
population in high redshift universe.
 Beckwith et al. (1998) argued that the  surface density
 of EROs with $R-K \ge 6$ mag and $K \le 19.75$ mag
 is 0.14 ${\rm arcmin}^{-2}$ for EROs with
 $R-K \ge 6$ and $K \le 19.75$.
Furthermore Thompson et al. (1999) derived
a surface density of EROs with  $R-K^{\prime}  \ge 6$ mag 
and $K^{\prime}  \le 19.0$ of 0.039 $\pm$ 0.016
 ${\rm arcmin}^{-2}$
and estimated that the volume density of bright EROs
to be as high as that of nearby Seyfert galaxies.
There are mainly two possible interpretations for EROs (We here
do not intend to discuss whether some EROs are actually
low-mass Galactic stars such as main-sequence M stars 
brown dwarfs).
One is that EROs are   passively evolving red elliptical galaxies
at high $z$ (e.g., those recently discovered
by the VLT; Ben\'\i tez et al. 1998), 
and the other is that EROs are dusty starburst
galaxies with  starburst components obscured heavily by dust
(e.g., HR10; Dey et al. 1999).
It still remains unclear  
which interpretation among the two
is more plausible,
essentially because
spectroscopic studies of EROs have not been so accumulated yet which
can reveal unambiguously the redshift of EROs
and thus discriminate the effects of aging of stellar populations
and those of dust extinction.

Recent observational results on EROs have begun to (or will soon begin to)
provide valuable information 
concerning this problem.
For example, Smail el al. (1999)
discovered two EROs
with $I-K \ge 6.0$ and 6.8 among  their SCUBA sources 
samples,
which implies that the submillimeter telescopes  with the improved
detectability (the detection limit of the order
of $10^2$ $\mu$Jy) can detect 
submillimeter flux indicative of the obscured
dusty starburst in a significant number of EROs. 
Furthermore statistical spectroscopic studies of EROs
by the already existing
large grand-based  telescopes (e.g., SUBARU with IRCS and FMOS)
will reveal the strength of $\rm H\alpha$ emission
that is not affected so strongly by dust extinction
and thereby clarify star formation rate and history
of EROs.
It is also possible
that the  future improved $HST$ ACS
and the 8m-class grand-based  telescopes
provide the detailed morphology of an ERO.
Accordingly, one of important works
for clarifying the origin of EROs
is to compare emission line strengths,
submillimeter flux, and morphological properties
observed (or those that will be observed in near future)  
in EROs 
with those predicted by a certain  theoretical dusty starburst model.
In order to make this comparison possible,
theoretical models of dusty starburst galaxies
should investigate the time evolution of
$R-K$ color, $K$ band magnitude,
emission lines, submillimeter flux, and morphology
for  a dusty starburst galaxy $in$ $a$ $self$-$consistent$ $manner$.
Although the present code still cannot predict emission and absorption
line strengths of a dusty starburst galaxy,
our new code enables us to investigate most of the above
important properties
simultaneously in  numerical simulations.
Thus it is  one of important issues for future numerical
and theoretical studies with the present new code
to clarify the origin of EROs.

\section{Conclusion}
We present a new code which is designed to derive a SED for 
an arbitrary spatial distribution of stellar and gaseous components
for a dusty starburst galaxy.
By using this code, we can calculate SEDs based on numerical simulations that
can analyze simultaneously dynamical and chemical evolution,
structural and kinematical properties, morphology, star formation history,
and transfer of metals and dust 
in interstellar medium for a starburst galaxy.
Accordingly, we can investigate variously different properties of starburst
galaxies, such as effects of dynamical evolution on galactic SEDs,
physical correlations between morphology and SEDs,
photometric evolution from UV to submillimeter  wavelength,
two dimensional distribution of $A_{\rm V}$, 
and dependence of SEDs on line-of-site of an observer.
Thanks to this code, we can furthermore try to clarify the origin
of possible  candidates of starburst galaxies, such as 
low $z$ ULIRGs (e.g., Sanders et al. 1988;  Sanders \& Mirabel 1996),
intermediate z ones (e.g., Tran et al. 1999),
faint SCUBA sources (Smail, Ivison, \& Blain 1997; 
 Barger et al. 1998;
Hughes et al. 1998; Smail et al. 1998; Ivison et al. 1999; Lilly et al.  1999),
EROs (Elston, Rieke, \& Rieke 1988; Dey et al. 1999; Smail  et al. 1999),
optically faint radii sources detected by VLA (e.g., Richards et al. 1999),
Lyman-break galaxies (Steidel et al. 1996; Lowenthal et al. 1997),
and emission line galaxies (e.g., Mannucci et al. 1998). 
By using a new  code developed in the present study, we try to answer the
following seven questions in our  forthcoming papers:
(1)  When and how a gas-rich major galaxy merger becomes
an ULIRG during the dynamical evolution of the merger ?
(2) How high z faint SCUBA sources with dusty starburst form and evolve?
(3) What are physical conditions for high z dusty galaxies to become EROs ? 
(4) Are there any evolutionary links between high z possible dusty
starburst galaxies, such as faint SCUBA sources, EROs, optically
faint radio sources recently detected by VLA, emission line galaxies,
and Lyman-break ones ?
(5) When does a forming disk galaxy show the strongest  submillimeter
flux? 
(6) What physical processes can determine the shapes of SEDs observed in
Lyman-break galaxies?
(7) How dusty interstellar gas affects apparent morphology of intermediate
and high $z$ dusty galaxies ?
Although this code has some disadvantages in deriving very precise 
SEDs of dusty starburst galaxies,
we believe that this code enables us to grasp some essential ingredients
of physical processes related to galaxy formation with starburst
at low and high z universe.

\acknowledgments
We are grateful for the referee G. D. Bothun for giving us
valuable comments and suggestions that greatly improve
our paper.
Y.S. thank to the Japan Society for Promotion of Science (JSPS) 
Research Fellowships for Young Scientist.

\clearpage


\figcaption{
Schematic representation of a new method for
estimating the effects of internal dust extinction
in a numerical simulation.
For simplicity, we here consider only one stellar particle
(filled circle) with metallicity $Z_{1}$ and SED $F_{sed,1}$
and four gaseous ones (open circles) with metallicity with $Z_{i}$ 
($i$ = 2 $\sim$ 5) and 
column density  ${N(\rm H)}_{i}$ ($i$ = 2 $\sim$ 5).
The description of $F_{sed,1}$ without abbreviation
is given in the main manuscript.
In the present study, ${N(\rm H)}_{i}$ is fixed at the same
value for all gas clouds.
This star is assumed to emit four rays ($N_{ray,1}$ = 4) that
are labeled as  $L_{1}$, $L_{2}$, $L_{3}$, and $L_{4}$.
Here each of rays has the SED  $F_{sed,1}$/4 and
dust extinction and emission are  calculated for each ray
according to the reddening formulation described in the main manuscript.
The ray $L_{1}$, $L_{2}$,  and $L_{4}$ penetrate
gas particles and accordingly the SED of each ray is corrected
by dust extinction according to the gaseous metallicity and
column density $Z_{i}$ and ${N(\rm H)}_{i}$ ($i$ = 2 $\sim$ 5).
In this figure, the region 
where the rays  $L_{1}$ and $L_{4}$ pass through
gas clouds are very different.
However, the SED of the ray $L_{1}$ after dust correction
is assumed to  be the same as that of the ray $L_{4}$ after dust correction
if  the metallicity $Z_{2}$ of the   gas particle penetrated by  $L_{1}$
is the same as  $Z_{4}$ of the  gas particle by $L_{4}$.
This is firstly
because we do not consider the detailed structure of gas clouds
and secondly because
column density is the same for all clouds in the present study:
Whichever region of a gas cloud
a ray penetrates,
the amount of dust absorption and reemission is the same for the ray.
The ray $L_{2}$ penetrates two gas clouds, therefore 
correction of the SED is done two times.
The ray $L_{3}$ does not penetrate gas particles at all,
and therefore the SED remains the same.
By using the above new method,
dust extinction ($A_{V}$) and SED
for each of stellar particles
and reemission of gaseous particles
can be estimated
in a simulation.
By summing stellar particles' SEDs corrected by dust 
and dust reemission from gaseous particles,
we derive a SED of a dusty galaxy. 
\label{fig-1}}

\figcaption{
Time evolution of mass distribution projected
onto $x$-$y$ plane (orbital plane) for dark halo (top),
stars (second from the top), gas (second from the bottom),
and new stars (bottom) in the standard merger model
 at each time  $T_{m}$.
The $T_{m}$ indicated in the upper
right-hand corner represents the time  that has elapsed
since the two disks begin to merge. 
Here the  scale is given in our units (17.5 kpc) and 
each of the  24  frames measures 200
kpc on a side.    
\label{fig-2}}

\figcaption{The same as Figure 2 but for mass distribution  projected onto
$x$-$z$ plane.
\label{fig-3}}

\figcaption{
Time evolution of star formation rate  
of the standard model. 
The star formation rate of the merger becomes maximum ($\sim 378 \rm  M_{\odot}/$yr)
at  $T_{m}$ = $1.3$ Gyr.
\label{fig-4}}

\figcaption{Mass distribution of stars (left),
gas (middle), and new star (right) at  $T_{m}$ = 1.13 Gyr 
projected for $x$-$y$ plane (upper) and $x$-$z$ one (lower)
in the standard model. 
Here the   scale is given in our units (17.5 kpc)  and 
each of the six frames measures
47 kpc on a side. 
This figure describes the mass distribution of the merger
at the pre-starburst epoch.
\label{fig-5}}

\figcaption{The same as Figure 5  but for  $T_{m}$ = 1.69 Gyr.
This figure describes the mass distribution of the merger
at the post-starburst epoch.
\label{fig-6}}

\figcaption{The spectral energy distributions (SEDs) of
the standard model with dust extinction  
at each time  $T_{m}$ (upper panel). 
Based on stellar and gaseous  distribution  
shown in Figure 1 and 2,  these SEDs are derived.
For comparison, the results of the model without dust extinction
are also given in the lower panel of this figure   for each time. 
The number  in the upper right  corner  of the lower panel indicates the
time $T_{m}$ in units of Gyr.  
\label{fig-7}}

\figcaption{
Mass distribution of the standard model projected
onto $x$-$y$ plane (upper four frames) and $x$-$z$ one (lower four ones)
 at $T_{m}$ = 1.3 Gyr corresponding
to the epoch of maximum starburst of the merger for total components (upper left),
old stellar components initially located in two disks (upper right), 
gaseous ones (lower left), and new stellar ones formed by secondary starburst (lower right). 
Each of the eight frames  measures 64  kpc on a side.
\label{fig-8}}

\figcaption{
The  rest-frame SED of the  galaxy merger at  
$T_{m}$ = 1.3 Gyr (black solid line) for $all$ stellar components.
For comparison, the SED of the merger without dust extinction and reemission
is also given by a red dotted line. 
Furthermore the SED for stellar component located
within the central 3.5 kpc is given by a green long-dashed line.
We can clearly see the effects 
of dust extinction and reemission on the SED shape
in the  merger at $T_{m}$ = 1.3 Gyr by comparing the black solid line 
and the red dotted one. 
For comparison,
the observed SED of Arp 220 by Rigopoulou, Lawrence, \& Rowan-Robinson (1996) 
is also given by a blue short-dashed  
line with blue open squares. The reason for our failure to
reproduce the observed SED around $10^{5} \rm \AA$  is essentially  that we do not
include the effects of small grains in the present study.
The SED within the central 3.5 kpc of the present
merger mode is more similar to the observed one.
\label{fig-9}}

\figcaption{
Two-dimensional distribution of $K$ band surface brightness 
projected onto $x$-$y$ plane at  $T_{m}$ = 1.3 Gyr in the standard model.
Each frame measures 64 kpc on a side and includes 400  bins (20 $\times$ 20).
The size of the frame is the same as that shown in Figure  8
and the scale is given in our units (17.5 kpc).
We here estimate the
mean values of surface brightness  for each bin based on
SEDs of stellar particles located within each bin. 
For the bin within  which no stellar particles are found to be located,
any color contours are not given  for clarity.
As is shown in the color legend of this figure,
the magnitude of the surface brightness ranges
from 16.1 in the central part to 26.6 mag ${\rm arcsec}^{-2}$ 
in the outer one. 
Note that the inner  region of the merger
is brighter in K-band owing to the central dusty starburst.  
\label{fig-10}}

\figcaption{The same as Figure 10 but for R-K color.
\label{fig-11}}

\figcaption{The same as Figure 10 but for $A_{V}$.
For clarity, color contours are described
only for the regions with $A_{V}$ larger than
0.1.  
\label{fig-12}}

\figcaption{Time evolution of $V$, $R$, 
and $K$ band magnitude
($M_{V}$, $M_{R}$, and $M_{K}$)
for the standard model with dust extinction (upper).
For comparison, the results of the model without dust extinction
are also shown in this figure (lower).
Note that owing to internal dust extinction,
the absolute magnitude of the merger with
dust extinction becomes smaller with the difference
between the two models depending on wavelength and time. 
\label{fig-13}}

\figcaption{The same as Figure 13 but for rest-frame  colors,
$V-I$, $I-K$, and $R-K$.
\label{fig-14}}

\figcaption{The same as Figure 13 but for 
total luminosity at 60 $\mu$m, 100 $\mu$m,
450 $\mu$m, and 850 $\mu$m. 
Note that the infrared flux becomes maximum
at the epoch of the maximum starburst ($T_{m}$ = 1.3 Gyr) in the merger.
\label{fig-15}}

\figcaption{The time evolution of dust temperature for
the model with dust extinction. 
\label{fig-16}}

\figcaption{
Redshift dependence  of $V-I$  in  the standard model
for  $T_{m}$ = 1.1 Gyr (pre-starburst phase), $T_{m}$ = 1.3 Gyr
(maximum starburst), and $T_{m}$ = 1.7 Gyr (post-starburst phase).
Here $V-I$  is the color in observed frame (not in rest-frame)
and z represents the redshift at which the two disks cause their maximum
starburst.
This figure describes how the $V-I$ color for
each phase of galaxy merging in observed frame
depends on the epoch of galaxy merging. 
\label{fig-17}}

\figcaption{
The same as Figure 17 but for $I-K$.
\label{fig-18}}

\figcaption{
The same as Figure 17 but for $R-K$.
\label{fig-19}}


\end{document}